\documentclass[aps,prd,floatfix,12pt]{revtex4}
\usepackage{epsfig}
\usepackage{bm}
\usepackage{dcolumn}
\usepackage{latexsym}

\begin{document}

\title{Lattice QED in an external magnetic field: Evidence for dynamical 
chiral symmetry breaking.}

\author{J.~B.~Kogut}
\affiliation{Department of Energy, Division of High Energy Physics, Washington,
DC 20585, USA \\
and \\
Department of Physics -- TQHN, University of Maryland, 82 Regents Drive,
College Park, MD 20742, USA}
\author{D.~K.~Sinclair}
\affiliation{HEP Division, Argonne National Laboratory, 9700 South Cass Avenue,
Lemont, IL 60439, USA}

\begin{abstract}
We simulate QED in a strong constant homogeneous external magnetic field on a
euclidean space-time lattice using the Rational Hybrid Monte Carlo method,
developed for simulating lattice QCD. Our primary goal is to measure the
chiral condensate in the limit when the input electron mass $m$ is zero. We
observe a non-zero value, indicating that the external magnetic field catalyzes
chiral symmetry breaking as predicted by approximate truncated Schwinger-Dyson
methods. Such behaviour is associated with dominance by the lowest Landau
level which causes the effective dimensional reduction from $3+1$~dimensions to
$1+1$ dimensions for charged particles (electrons and positrons) where the
attractive forces of QED can produce chiral symmetry breaking with a dynamical
electron mass and associated chiral condensate. Since our lattice simulations
use bare (lattice) parameters, while the Schwinger-Dyson analyses work with
renormalized quantities, direct numerical comparison will require 
renormalization of our lattice results.

\end{abstract}

\maketitle

\section{Introduction}
Theoretical studies of free electrons in external electromagnetic fields were
some of the earliest applications of relativistic quantum mechanics, 
\cite{Sauter,Heisenberg,Schwinger:1951nm}, and
exhibited some of the features of quantum electrodynamics, such as the 
Sauter-Schwinger effect \cite{Sauter,Schwinger:1951nm}
(the instability of the vacuum in a strong electric field,
with the production of electron-positron pairs). [For a review of those field
configurations where an exact closed form solution as simple integral or 
series is known, see, for example, the review article by Gerald Dunn.
\cite{Dunne:2012vv}]
 
QED (quantum electrodynamics) in strong external electromagnetic fields is of 
relevance to high and medium energy physics, laser and accelerator physics, 
astrophysics and cosmology, and condensed matter physics. See, for example, 
V.~Yakimenko {\it et. al.} \cite{Yakimenko:2018kih},
for a discussion with examples of the physics 
associated with QED in strong electromagnetic fields.

Current and future experiments at SLAC, LBNL and ELI, which collide the light 
from petawatt lasers with electron/positron beams or plasmas, produce 
environments with electromagnetic fields strong enough to produce quantum and 
QED effects \cite{Fedeli:2020fwt,Meuren:2021qyv,Yu:2023}. 
This requires electromagnetic fields which approach or exceed the
critical values $E_{cr}=m^2/e$ or $B_{cr}=m^2/e$ where $m$ and $e$ are the 
electron mass and charge respectively. Another source of such strong fields is
in the charged particle beams at future accelerators and their interactions,
where fields could even be strong enough that the electron loop expansion 
(the last remnant of perturbation theory) breaks down \cite{Yakimenko:2018kih}.
Certain compact 
astronomical objects, in particular those which are sources of X- and gamma-ray
emissions and are identified as neutron stars are believed to have very strong
surface magnetic fields (magnetars). See for example\cite{Harding:2006qn}. 
Some have postulated the presence of very
strong magnetic fields in the early universe as the source of the magnetic
fields observed in the universe today.

We simulate lattice QED in external electromagnetic fields using the methods 
developed for simulating lattice QCD. Although some of the more interesting
physics from an experimental point of view such as the Sauter-Schwinger effect
-- the production of electron-positron pairs from the unstable vacuum -- are
produced by strong external electric fields, this electric field makes the
action complex, the imaginary part describing vacuum decay. Hence standard
lattice simulations cannot be applied. We therefore start with simulations
of QED in external magnetic fields where the action is real and bounded below.
We simulate using a non-compact gauge action and staggered fermions, using the
RHMC algorithm to allow tuning to a single electron `flavour'. For details see
section~2.

We start with considering the case of `free' electrons in a constant (in 
space and time) magnetic field with magnitude $B$, comparing our lattice
results with the known exact solutions \cite{Schwinger:1951nm,Dunne:2012vv}
to determine the range of $B$ or more precisely $eB$ ($e$ is the electron 
charge) over which there is good agreement.
 
    Next we perform simulations of full lattice QED at a near-physical electron
charge $\alpha=e^2/4\pi=1/137$, on a $36^4$ lattice with safe ($36\,m >> 1$,
$m << 1$) electron masses $m=0.1$, $m=0.2$, comparing observables with those 
for free ($\alpha=0$) electrons in a magnetic field. We then perform 
simulations at $B=0$ for a range of $\alpha$ on $36^4$ lattices with $m=0.1$,
checking that the gauge action per site's $\alpha$ dependence is consistent 
with perturbation theory.

    One of the theoretically most interesting predictions for QED in an
external magnetic field is that the presence of this external field breaks
chiral symmetry at $m=0$. This manifests itself by giving a small dynamical
mass to the electron which in turn gives rise to a non-zero chiral condensate.
These predictions were obtained using truncated Schwinger-Dyson analyses, where
the effects of the truncations required to obtain results, are difficult to
estimate. The dynamical mass production was predicted in
\cite{Gusynin:1994xp,Gusynin:1995nb,Gusynin:1999pq,Gusynin:2000tv,
Leung:2005yq,Leung:2005xz,Leung:1996poh,Alexandre:2000nz,Alexandre:2001vu,
Wang:2007sn,Hattori:2017qio}.
Estimates of the chiral condensates are given in
\cite{Shushpanov:1997sf,Lee:1997zj}
For a good review with a more complete set of references see 
\cite{Miransky:2015ava}. For a more recent review, see, for example
\cite{Hattori:2023egw}. 
It is therefore important to check these predictions using methods
whose errors are easier to estimate and which allow systematic improvements.
Lattice QED simulations are of this nature. However, the best estimates of
Gusynin, Miransky and Shovkovy \cite{Gusynin:1999pq} 
predict a dynamical electron mass at our chosen
$eB=0.4848...$ and a near-physical $\alpha=1/137$, 
$m_{dyn}\sim 3 \times 10^{-35}$, far below anything we could measure on the
lattice. Therefore we choose a stronger electron charge $\alpha=1/5$, where the
predicted {$m_{dyn} \approx 3\times 10^{-4}$. Here our simulations show 
evidence of chiral-symmetry breaking at a level 1 to 2 orders of magnitude
greater than the `best' Schwinger-Dyson results, however, our lattice QED
results are for bare quantities, whereas those using Schwinger-Dyson analyses 
are for renormalized quantities. 

Preliminary results were presented at Lattice~2021, Lattice~2022 and 
Lattice~2023 \cite{Sinclair:2021plv,Sinclair:2022ykg,Sinclair}

Section~2 defines Lattice QED in an external magnetic field.
Section~3 compares the effects of an external magnetic field on electrons on
the lattice with those in the continuum.
Section~4 presents simulations of Lattice QED at $\alpha=1/137$ in an external
magnetic field.
Section~5 describes simulations of Lattice QED at $\alpha=1/5$ in an external 
magnetic field.
Section~6 presents discussions and conclusions.  

\section{Lattice QED in an external Magnetic Field}}

In this section we describe the lattice transcription of QED in an external
magnetic field used for our simulations. We use the non-compact gauge action
\begin{equation}
S(A) = \frac{\beta}{2}\sum_{n,\mu < \nu}[A_\nu(n+\hat{\mu})-A_\nu(n)
                                        -A_\mu(n+\hat{\nu})+A_\mu(n)]^2
\end{equation}
where $n$ is summed over the lattice sites and $\mu$ and $\nu$ run from $1$ to
$4$ subject to the restriction. $\beta=1/e^2$. The functional integral to
calculate the expectation value for an observable ${\cal O}(A)$ is then
\begin{equation}
\langle{\cal O}\rangle = \frac{1}{Z}\int_{-\infty}^\infty \Pi_{n,\mu}
           dA_\mu(n) e^{-S(A)}[\det{\cal M}(A+A_{ext})]^{1/8}{\cal O}(A)
\end{equation}
where ${\cal M} = M^\dag M$ and $M$ is the staggered fermion action in the
presence of the dynamic photon field $A$ and external photon field $A_{ext}$
describing the magnetic \cite{Leung:1996poh} field $B$ (or rather $eB$). $M$ 
is defined by 
\begin{equation}
M(A+A_{ext}) = \sum_\mu D_\mu(A+A_{ext})+m
\end{equation}
where the operator $D_\mu$ is defined by
\begin{eqnarray}
[D_\mu(A+A_{ext})\psi](n) &=&
\frac{1}{2}\eta_\mu(n)\{e^{i(A_\mu(n)+A_{ext,\mu}(n))}\psi(n+\hat{\mu})
\nonumber\\ 
&-&e^{-i(A_\mu(n-\hat{\mu})+A_{ext,\mu}(n-\hat{\mu}))}\psi(n-\hat{\mu})\} 
\nonumber\\
\end{eqnarray}
and $\eta_\mu$ are the staggered phases. Note that this treatment of the
gauge-field--fermion interactions is compact and so has period $2\pi$ in the
gauge fields.

We implement the RHMC simulation method of Clark and Kennedy 
\cite{Clark:2006fx}, 
using a rational approximation to ${\cal M}^{-1/8}$ and rational 
approximations to 
${\cal M}^{\pm 1/16}$. To account for the range of normal modes of the 
non-compact gauge action, we vary the trajectory lengths $\tau$ over the 
range,
\begin{equation}
\frac{\pi}{2\sqrt{\beta}} \leq \tau \leq
\frac{4\pi}{\sqrt{2\beta(4-\sum_\mu\cos(2\pi/N_\mu))}},
\end{equation}
of the periods of the modes of this gauge action. Here $N_\mu$ is the length
of the lattice in the $\mu$ direction which is chosen to be that of maximum
extent \cite{Hands:1992uv}. (Note that we typically only change the trajectory
length after a trajectory has been accepted.)

$A_{ext}$ are chosen in the symmetric gauge as
\begin{eqnarray}
 A_{ext,1}(i,j,k,l)=        -\frac{eB}{2}&&(j-1) \;\;\;\; i \ne N_1 \nonumber\\
 A_{ext,1}(i,j,k,l)= -\frac{eB}{2}(N_1+1)&&(j-1) \;\;\;\; i  =  N_1 \nonumber\\
 A_{ext,2}(i,j,k,l)=        +\frac{eB}{2}&&(i-1) \;\;\;\; j \ne N_2 \nonumber\\
 A_{ext,2}(i,j,k,l)= +\frac{eB}{2}(N_2+1)&&(i-1) \;\;\;\; j  =  N_2 \nonumber\\
\end{eqnarray}
while $A_{ext,3}(n)=A_{ext,4}(n)=0$ \cite{Alexandre:2001pa}. 
In practice we subtract the average 
values of $A_{ext},\mu$ from these definitions. This choice produces a 
magnetic field $eB$ in the $+z$ direction on every $1,2$ plaquette except that
with $i=N_1$, $j=N_2$, which has the magnetic field $eB(1-N_1N_2)$. Because 
of the compact nature of the interaction, requiring $eB N_1 N_2 = 2\pi n$ for
some integer $n=0,1,......N_1N_2/2$ makes the value of this plaquette 
indistinguishable from $eB$. Hence $eB=2\pi n/(N_1N_2)$ lies in the interval 
$[0,\pi]$.

At the end of each accepted trajectory, we subtract the multiple of 
$2\pi/N_\mu$ from each $A_\mu$ which reduces the magnitude of the lattice 
average of said $A_\mu$ to lie in the range $(-\pi/N_\mu,\pi/N_\mu]$. Since 
this is a gauge transformation it does not change any physics. In addition we
transform to Landau gauge. Both these operations aim to prevent the gauge 
fields from becoming too large.

One of the observables we calculate is the electron contribution to the
effective action per site $\frac{-1}{8V}{\rm trace}[\ln({\cal M})]$. For
$\ln({\cal M})$ we use a $(30,30)$ rational approximation to the logarithm.
Here we use the Chebyshev method of Kelisky and Rivlin \cite{KR}. 
While this has worse errors than a Remez approach, it preserves some of the 
properties of the logarithm itself, and is applicable on the whole complex 
plane cut along the negative real axis. (This would be important if we had an
electric field, in which case ${\det{\cal M}}$ would be complex).

\section{`Free' electrons in an external magnetic field}

We restrict ourselves to considering an external magnetic field of strength 
$B$ which is constant in space and time and oriented in the $+z$ direction. 
Classically `free' electrons in such a magnetic field traverse helical orbits.
The motion parallel to the magnetic field is free, while that orthogonal to 
the magnetic field is circular and hence bound.

Quantum mechanics restricts the motion perpendicular to the field to a discrete
set of transverse energy levels known as the Landau levels \cite{Akhiezer}.
As $B$ increases, the radii of these orbits decreases. The radius of the 
lowest Landau level is $1/\sqrt{eB}$. This leads to an effective dimensional 
reduction from $3+1$ to $1+1$ dimensions for charged particles in a large 
magnetic field. The energy of the n-th Landau level is:
\begin{equation}
E_n = \sqrt{p_3^2+2eBn+m^2}
\end{equation}
where $n=0,1,2...$. The degeneracy of each $n=0$ energy level is 
$\frac{eB}{2\pi}{\cal A}_{xy}$ and that of each $n > 0$ energy level is 
$\frac{eB}{\pi}{\cal A}_{xy}$ where ${\cal A}_{xy}$ is the area in the $xy$
plane.

We now calculate the chiral condensate $\langle\bar{\psi}\psi\rangle$ on the 
lattice for masses $m=0.1$ and $m=0.2$ on a $36^4$ lattice, values which we 
use for our initial QED simulations at $\alpha=1/137$, over a range of allowed
$eB$ values. This is to test the range of applicability of the lattice approach
against the known continuum results. The lattice chiral condensate
\begin{equation}
\langle\bar{\psi}\psi\rangle = \frac{1}{4V}{\rm trace}[M^{-1}(A_{ext})]
\label{eqn:pbp0}
\end{equation}
where $M$ is defined in section~2. Because $A_{ext}$ is independent of $z$ and
$t$, we only need to calculate the trace over one $xy$ plane. In fact, for 
$m=0.1$ and $m=0.2$, all terms in the trace are almost identical. We compare
this with the known continuum result:
\begin{equation}
\langle\bar{\psi}\psi\rangle=\langle\bar{\psi}\psi\rangle|_{eB=0}
+\frac{meB}{4\pi^2}\int_0^\infty \frac{ds}{s}e^{-sm^2}\left[\coth(eBs)
  -\frac{1}{eBs}\right].
\end{equation}
where $\langle\bar{\psi}\psi\rangle|_{eB=0}$ is taken from the lattice. This
is necessary because $\langle\bar{\psi}\psi\rangle|_{eB=0}$ depends on the UV
regulator, which is different on the lattice than in the continuum. (Note that
we have checked that we get the same result from equation \ref{eqn:pbp0} and 
from summing over the normal modes on the lattice.)

\begin{figure}[htb]
\epsfxsize=4.0in
\epsffile{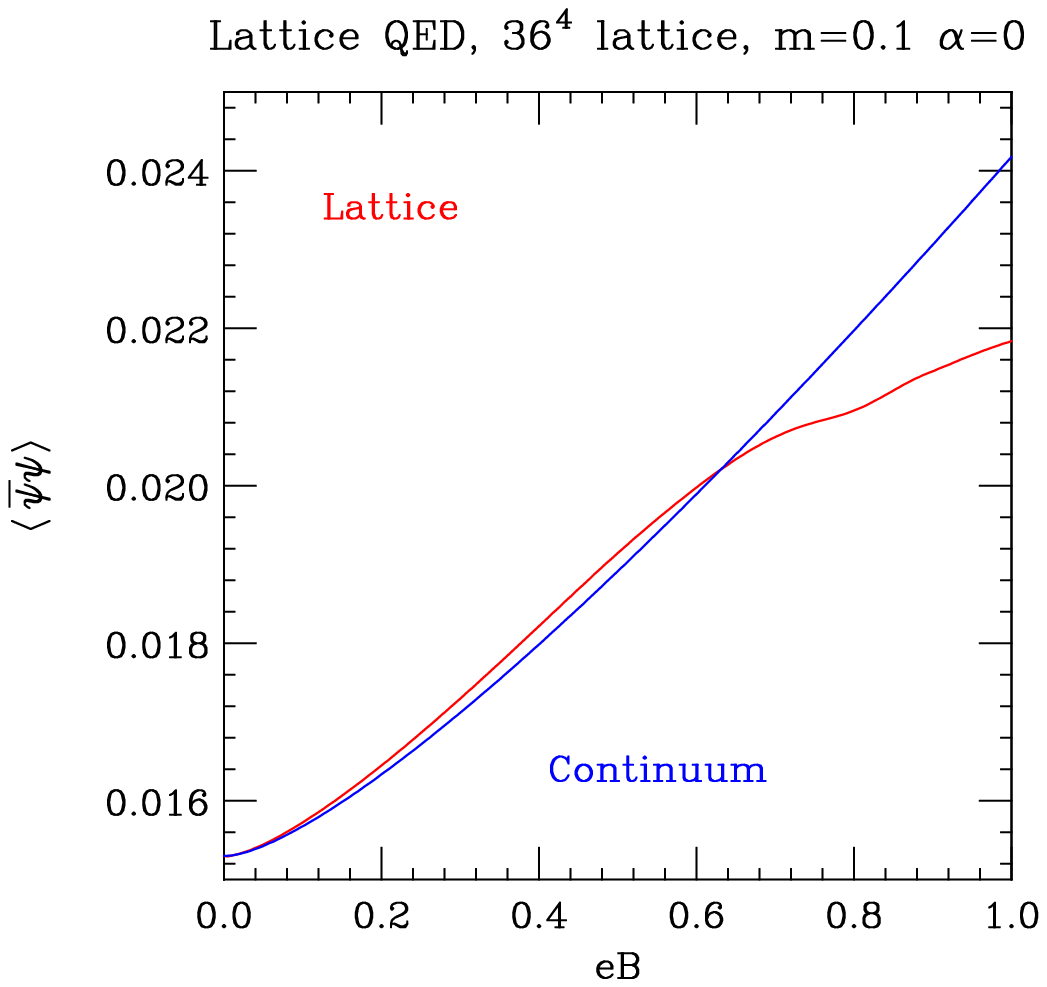}
\centerline{     }
\epsfxsize=4.0in
\epsffile{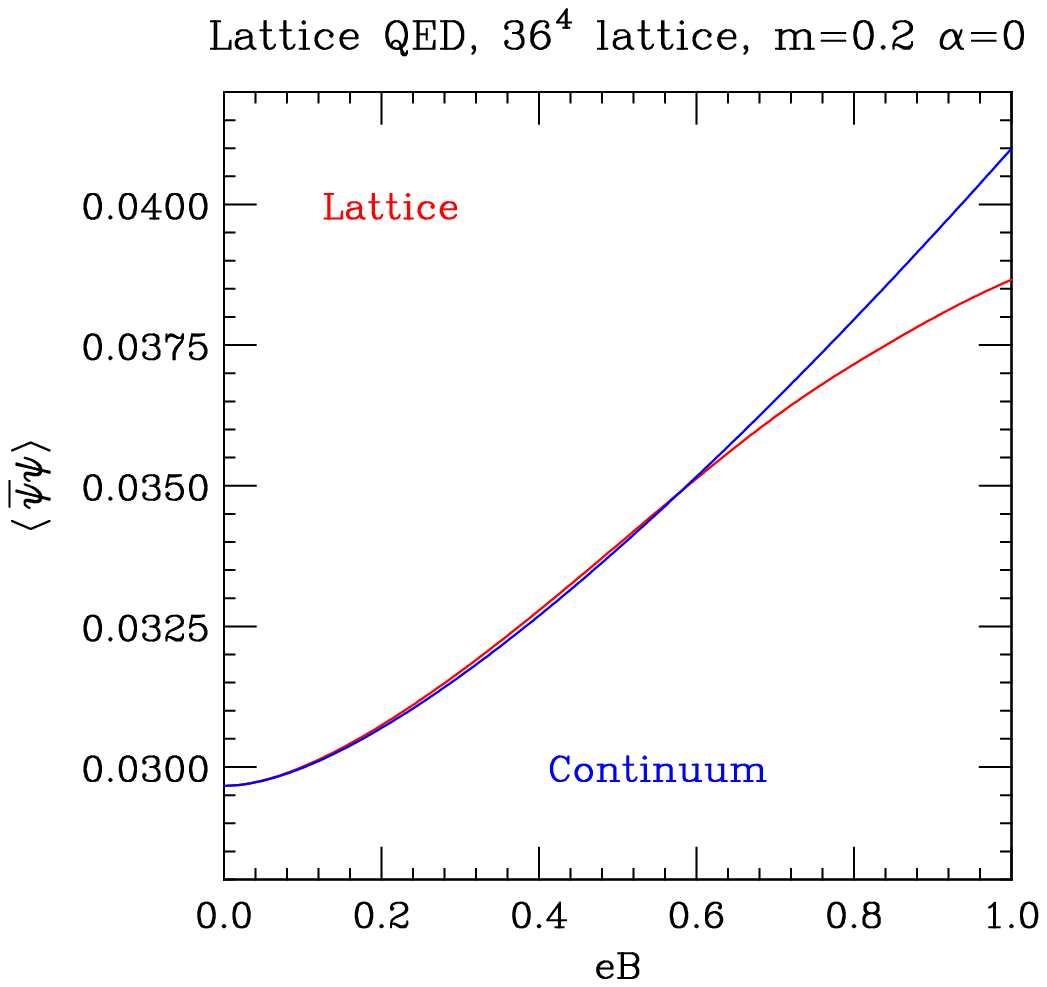}
\caption{Electron chiral condensates $\langle\bar{\psi}\psi\rangle$ as 
functions of $eB$, comparing the continuum and lattice results for a) $m=0.1$ 
and b) $m=0.2$.} 
\label{fig:pbp_free}
\end{figure}

Figure~\ref{fig:pbp_free} compares the chiral condensates for free electrons
in an external magnetic field $B$ for $m=0.1$ and for $m=0.2$ as functions of
$eB$ on a $36^4$ lattice, and compares them with the known continuum results.
We conclude that the lattice results are in acceptable agreement with the
continuum results for $eB\lesssim0.63$ for each mass.

We also calculate the fermion effective action/site on the lattice
\newpage
\begin{eqnarray}
  {\cal L}_f &=& -\frac{1}{4V}\ln\{\det[M(A_{ext})]\}        \nonumber    \\
             &=& -\frac{1}{4V}{\rm trace}\{\ln[M(A_{ext})]\} \nonumber    \\
\end{eqnarray}
which we compare with the known continuum result:
\begin{eqnarray}
{\cal L}_f &=& {\cal L}_f|_{B=0}
+\frac{(eB)^2}{24\pi^2}\int_0^\infty\frac{ds}{s}e^{-m^2s} \nonumber\\
&+&\frac{eB}{8\pi^2}\int_0^\infty\frac{ds}{s^2}e^{-m^2s}\left[\coth(eBs)
  -\frac{1}{eBs}-\frac{eBs}{3}\right]. \nonumber \\
\end{eqnarray}
Again we replace the divergent part of this quantity in the continuum version 
with that of the lattice version to take into account the difference between 
the continuum and lattice regulators. Since the quadratically divergent part 
of ${\cal L}_f$, ${\cal L}_f|_{eB=0}$ is a constant (independent of $eB$) 
and the logarithmically divergent term is proportional to $(eB)^2$, while the 
leading contribution of the integral (the finite part) is of order $(eB)^4$, 
the separation of the divergent and the finite parts is straight forward in 
principle.

\begin{figure}[htb]
\epsfxsize=4.0in
\epsffile{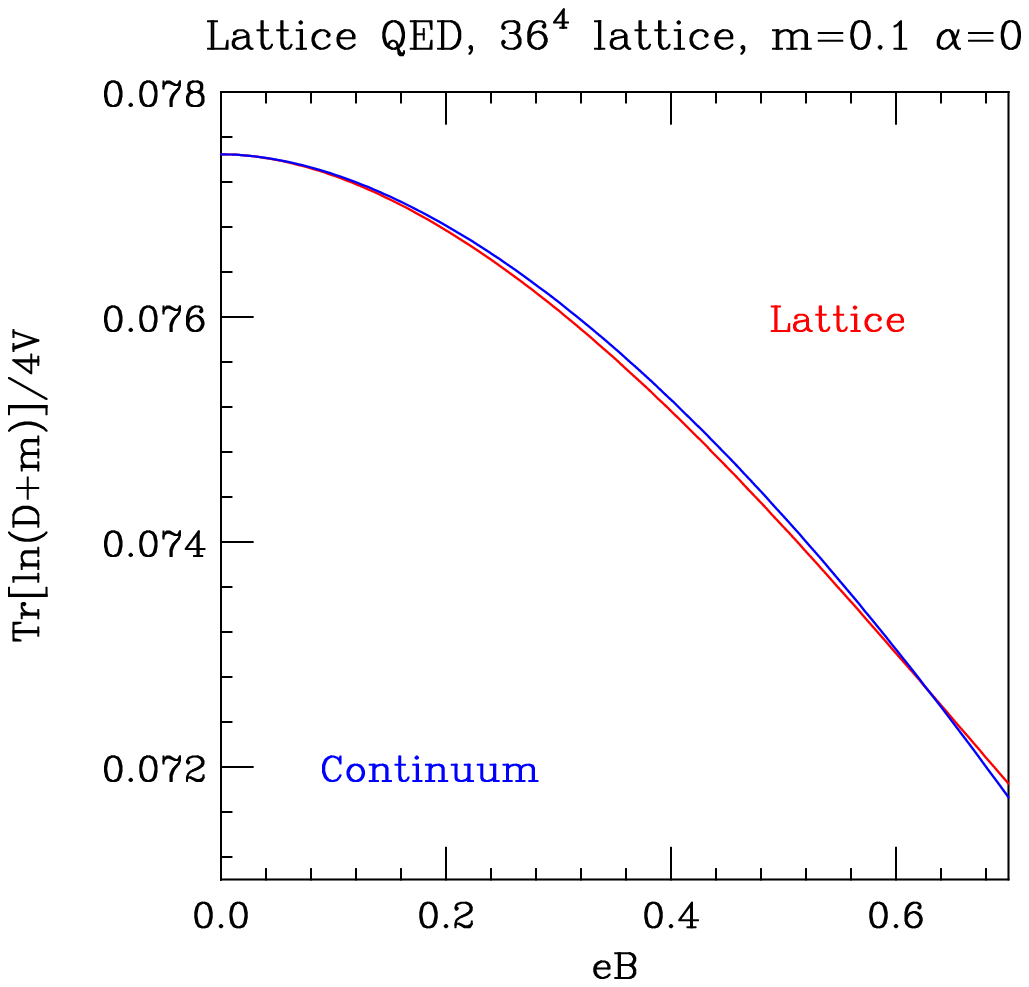}
\centerline{     }
\epsfxsize=4.0in
\epsffile{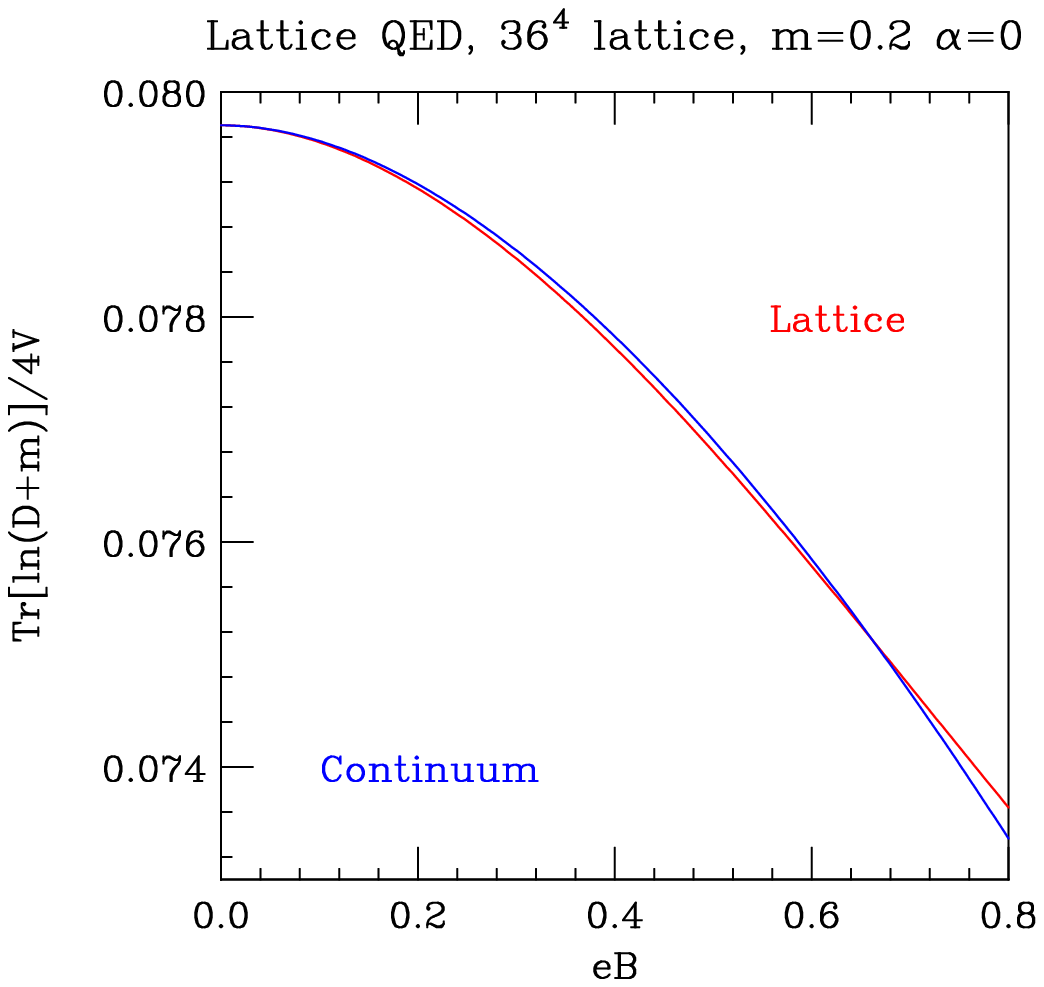}
\caption{-Electron effective actions/site $-{\cal L}_f$ as functions of $eB$, 
comparing the continuum and lattice results for a) $m=0.1$ and b) $m=0.2$.}
\label{fig:logD_free}
\end{figure}

Figure~\ref{fig:logD_free} shows the fermion contributions to the effective 
action on a $36^4$ lattice as functions of $eB$, comparing the lattice and
continuum results for $m=0.1$ and $m=0.2$. Again, we see good agreement over
a range of $eB$ values at least as large as for the chiral condensate.

\clearpage

\section{Lattice QED simulations at $\alpha=1/137$}

We simulate lattice QED on a $36^4$ lattice using the approach presented in 
section~2 at lattice (bare) coupling $\alpha=1/137$ and lattice (bare)  
masses $m=0.1$ and $m=0.2$. At this $\alpha$, the difference between bare and 
renormalized parameters is small (at most a few percent), and will therefore 
be ignored. Hence we consider these simulations to be performed at the 
physical electron charge. For $m=0.1$ we simulate over a range of allowed 
$eB$ values in the interval $0 \le eB \le 2\pi\times160/36^2=0.7757...$, while
for $m=0.2$ we use a selection of allowed $eB$ values in the range
$0 \le eB \le 2\pi\times200=0.9696...$. At each $eB$ we run for 12500
trajectories of lengths randomly chosen from the periods of the modes of the
gauge action. We store a gauge configuration every 100 trajectories for future
analyses. Note that, since QED probably does not have a UV completion, the
action we choose should be considered to define an effective field theory. 
Its form is chosen to generate results consistent with QED perturbation theory
with a lattice regulator.  

\begin{figure}[htb]
\epsfxsize=4.0in
\epsffile{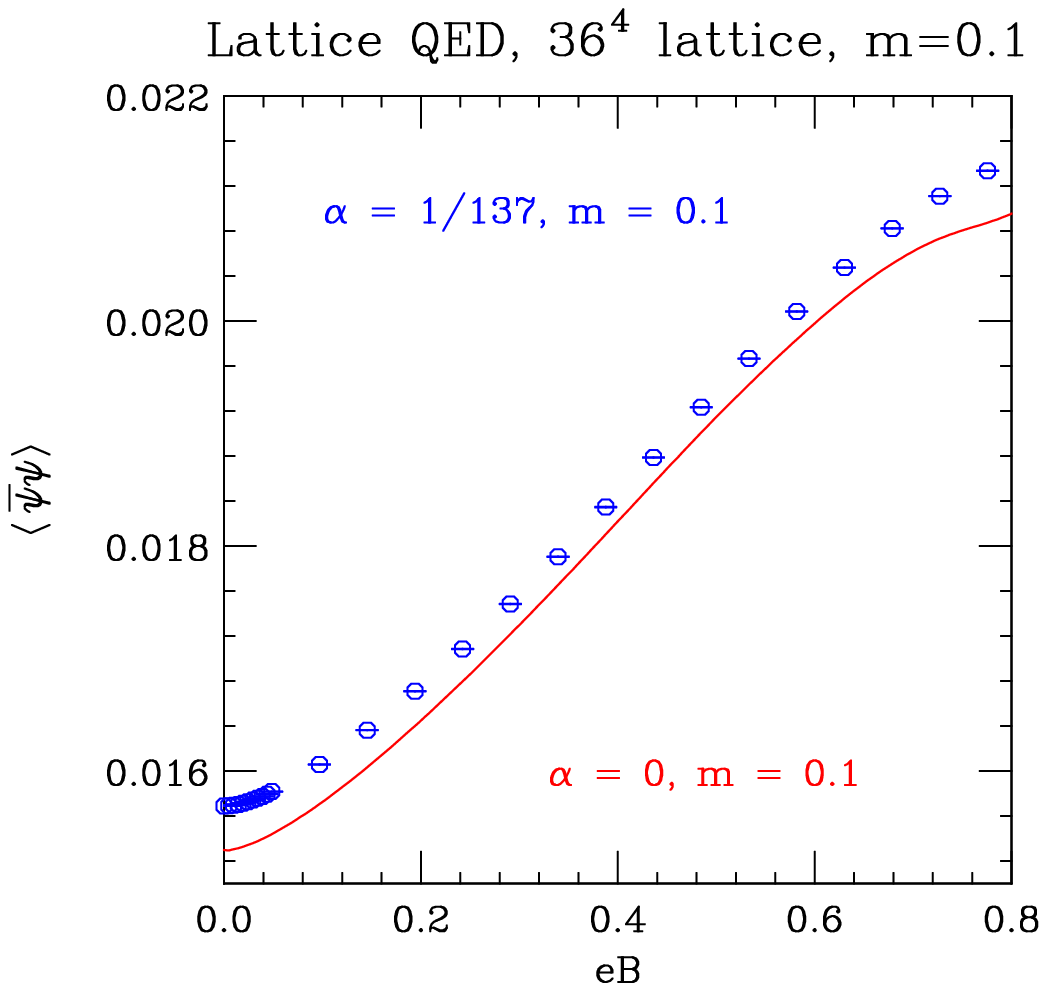}
\centerline{     }
\epsfxsize=4.0in
\epsffile{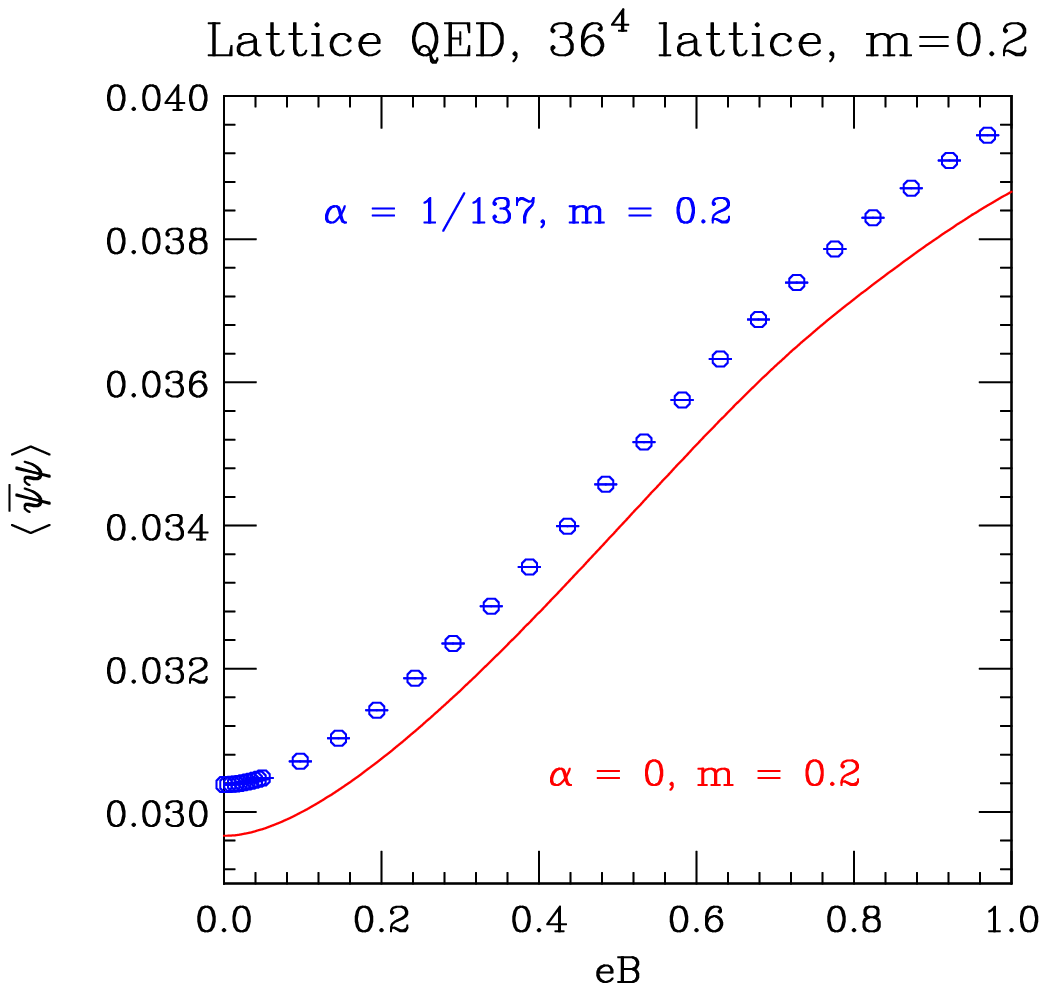}
\caption{Electron chiral condensates $\langle\bar{\psi}\psi\rangle$ as
functions of $eB$, comparing the $\alpha=1/137$ lattice results with the 
free-field ($\alpha=0$) lattice results for a) $m=0.1$ and b) $m=0.2$.}
\label{fig:pbp_1/137}
\end{figure}

In figure~\ref{fig:pbp_1/137} the chiral condensates as functions of $eB$ for 
$\alpha=1/137$ are compared to those for free electrons in a magnetic field
($\alpha=0$). In both cases the condensate for $\alpha=1/137$ lies above that
for free electrons. This was to be expected, since the attractive force between
electrons and positrons in QED is predicted to enhance the chiral condensate.
A truncated Schwinger-Dyson approach indicates that chiral symmetry breaking
with a dynamical electron mass proportional to $\sqrt{eB}$ and a non-zero 
chiral condensate proportional to $(eB)^{3/2}$ survives in the $m\rightarrow0$ 
limit, however small $\alpha$ might be. However, at $\alpha=1/137$, as stated
in the introduction, the dynamical electron contribution to the electron mass
over the range of $eB$s accessible to these simulations is negligable so that
these measurements could be checked by lattice perturbation theory.

\begin{figure}[htb]
\epsfxsize=4.0in
\epsffile{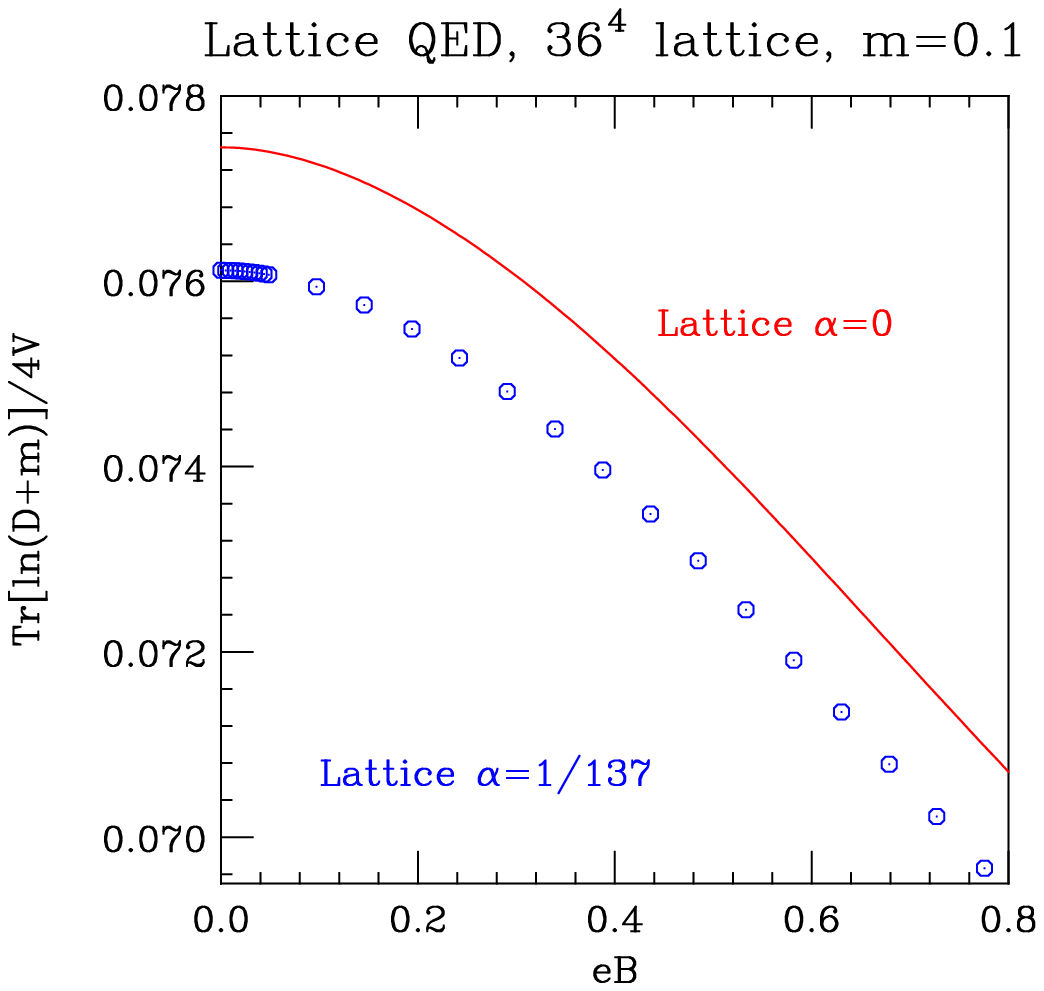}
\centerline{     }
\epsfxsize=4.0in
\epsffile{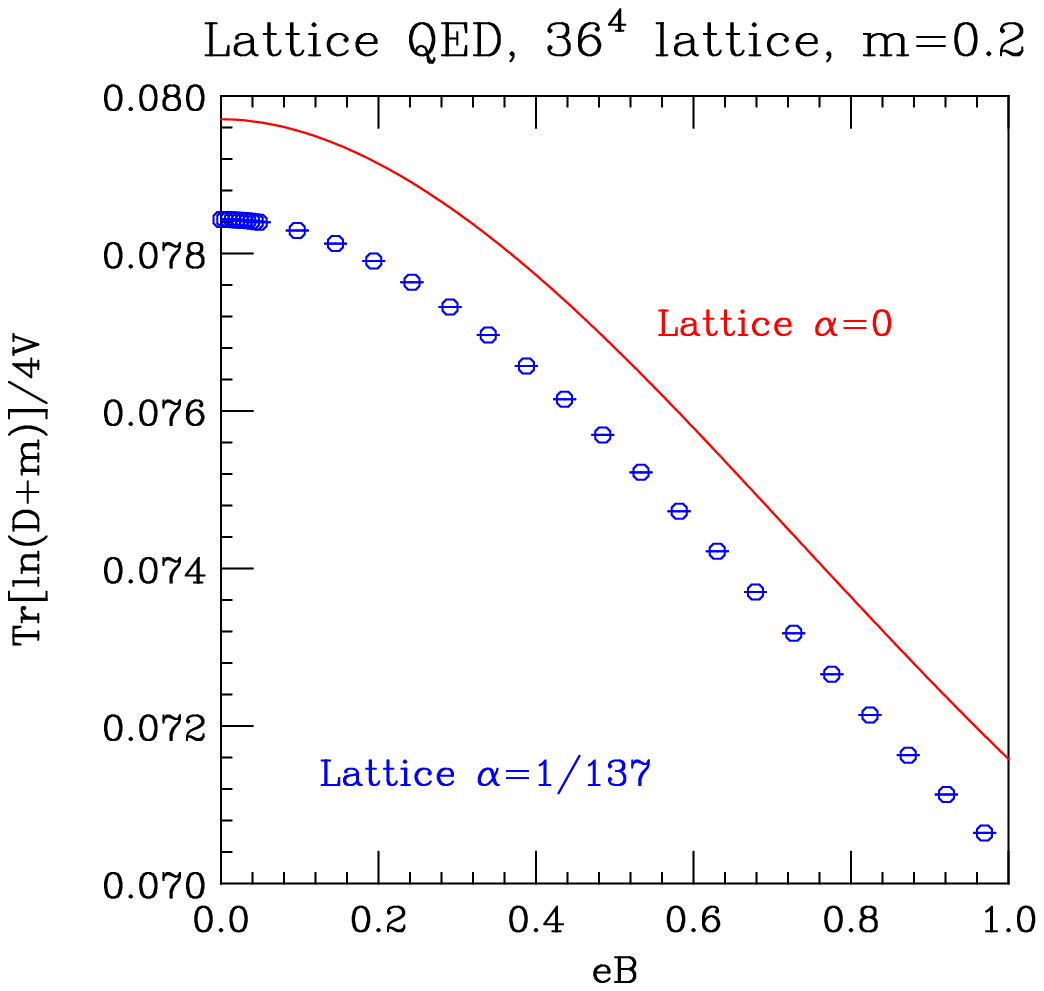}
\caption{-Electron effective actions/site $-{\cal L}_f$ as functions of $eB$,
comparing the $\alpha=1/137$ lattice results with the free-field ($\alpha=0$)
lattice results for a) $m=0.1$ and b) $m=0.2$.}
\label{fig:logD_1/137}
\end{figure}

Figure~\ref{fig:logD_1/137} compares the effective fermion actions/site 
\begin{eqnarray}
  {\cal L}_f &=& -\frac{1}{4V}\ln\{\det[M(A+A_{ext})]\}        \nonumber  \\
             &=& -\frac{1}{4V}{\rm trace}\{\ln[M(A+A_{ext})]\} \nonumber  \\
\end{eqnarray}
for QED with $\alpha=1/137$, to its free-electron value as functions of $eB$
for $m=0.1$ and $m=0.2$. The effective actions for QED are well above those 
without QED.

\clearpage

\section{Lattice QED simulations at $\alpha=1/5$, $eB=2\pi\times100/36^2$}

We are interested in finding evidence of chiral symmetry breaking for QED in 
(strong) external magnetic fields, which survives in the $m \rightarrow 0$
limit. Since Schwinger-Dyson analyses predict that this manifests itself as a 
dynamical electron mass, which at $\alpha=1/137$ and our chosen magnetic field
$eB=0.4848...$ is $m_{dyn}\approx 3 \times 10^{-35}$, in the $m=0$ limit, far 
below anything we could measure on the lattice, we simulate at a far stronger 
bare coupling, $\alpha=1/5$. For a renormalized $\alpha=1/5$, the predicted 
$m_{dyn} \approx 3\times 10^{-4}$ \cite{Gusynin:1999pq} equation (51). (Note 
that \cite{Leung:2005yq} equation (4.22) gives essentially the same result at
renormalized $\alpha=1/5$).  This leads to 
$\langle\bar{\psi}\psi\rangle \approx 1.2 \times 10^{-4}$, \cite{Lee:1997zj} 
equation (B4), which should be measurable. Of course, since our $\alpha=1/5$ is
the bare(lattice) $\alpha$, there is no guarantee that it will give a 
measurable result, we only know this a postiori.

First, we need to determine that $\alpha=1/5$ is still in the range of 
$\alpha$ values where perturbation theory holds. We do this by performing 
simulations over a range of $\alpha$ values from $\alpha=0$ to $\alpha=1/5$. 
These simulations were performed at $eB=0$ and $m=0.1$ Our chosen observable 
for performing this test is the gauge Lagrangian ${\cal L}_\gamma$, the gauge 
action/lattice site. For $\alpha=0$, ${\cal L}_\gamma=1.5$ by the equipartition
theorem. This value changes (decreases) only slowly with increasing $\alpha$.

\begin{figure}[htb]
\epsfxsize=4.0in
\epsffile{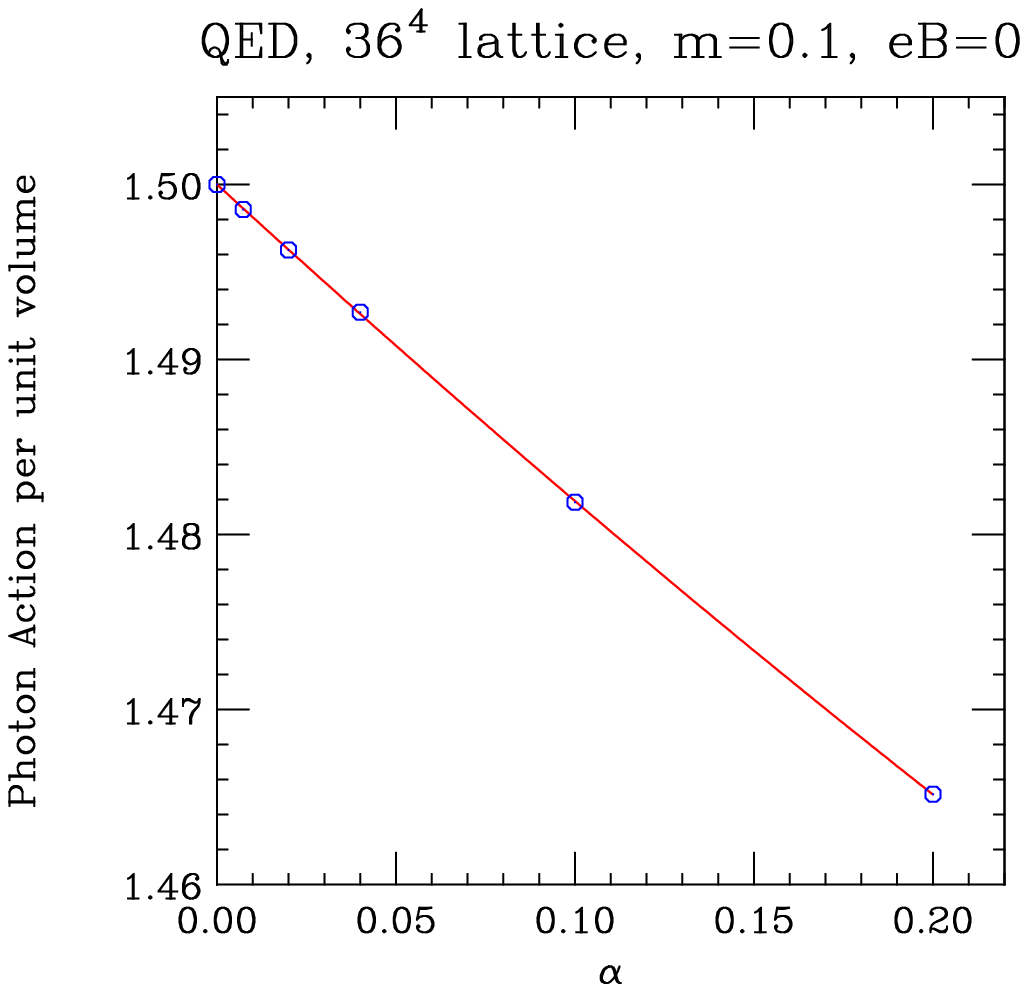}
\caption{Photon Lagrangian density as a function of $\alpha$ for $eB=0$. The 
curve is the fit \newline
${\cal L}_\gamma(\alpha)=1.5 - 0.187499 \alpha + 0.0661374 \alpha^2$.}
\label{fig:L_gamma}
\end{figure} 

The fact that ${\cal L}_\gamma$ is well approximated by a low order polynomial
in $\alpha$ with coefficients of decreasing magnitude over the range 
$0 \leq \alpha \leq 0.2$ is consistent with the claim that $\alpha=0.2$ lies
in the perturbative domain for $eB=0$. We have also performed simulations at
$\alpha=1$, m=0.1 for which ${\cal L}_\gamma(1)=1.49047(3)$, that does not
appear to be consistent with perturbation theory, although one would need
more simulations in the range $0.2 \leq \alpha\leq 1.0$ to check this. 

Having chosen $\alpha=1/5=0.2$ for our lattice QED simulations at strong 
coupling, we first perform simulations at $eB=0$. Since we do not expect any
chiral symmetry breaking breaking in the $m\rightarrow0$ limit, the chiral
condensate at non-zero mass should be proportional to $m\Lambda^2$  
probably with logarithmic corrections, where $\Lambda$ is the momentum cutoff.
On the lattice, since we use units where the lattice spacing is $1$, $\Lambda=\pm\pi$ in each of the 4 directions $x,y,z,t$. This means that the trace of the
propagator, which defines the chiral condensate, is expected to be dominated by
large momenta and hence short distances, and should not be sensitive to the 
size of the lattice. Therefore we should be able to calculate the condensate on
relatively small lattices, even in the limit as $m\rightarrow0$. We simulate 
over a range of electron masses $0.001 \leq m \leq 0.2$ on a $36^4$ lattice, 
noting that for the lowest mass $m=0.001$, $m N_\mu = 0.036 << 1$, far outside
the range of `safe' values for which would require $m N_\mu >> 1$.

\begin{figure}[htb]
\epsfxsize=4.0in
\epsffile{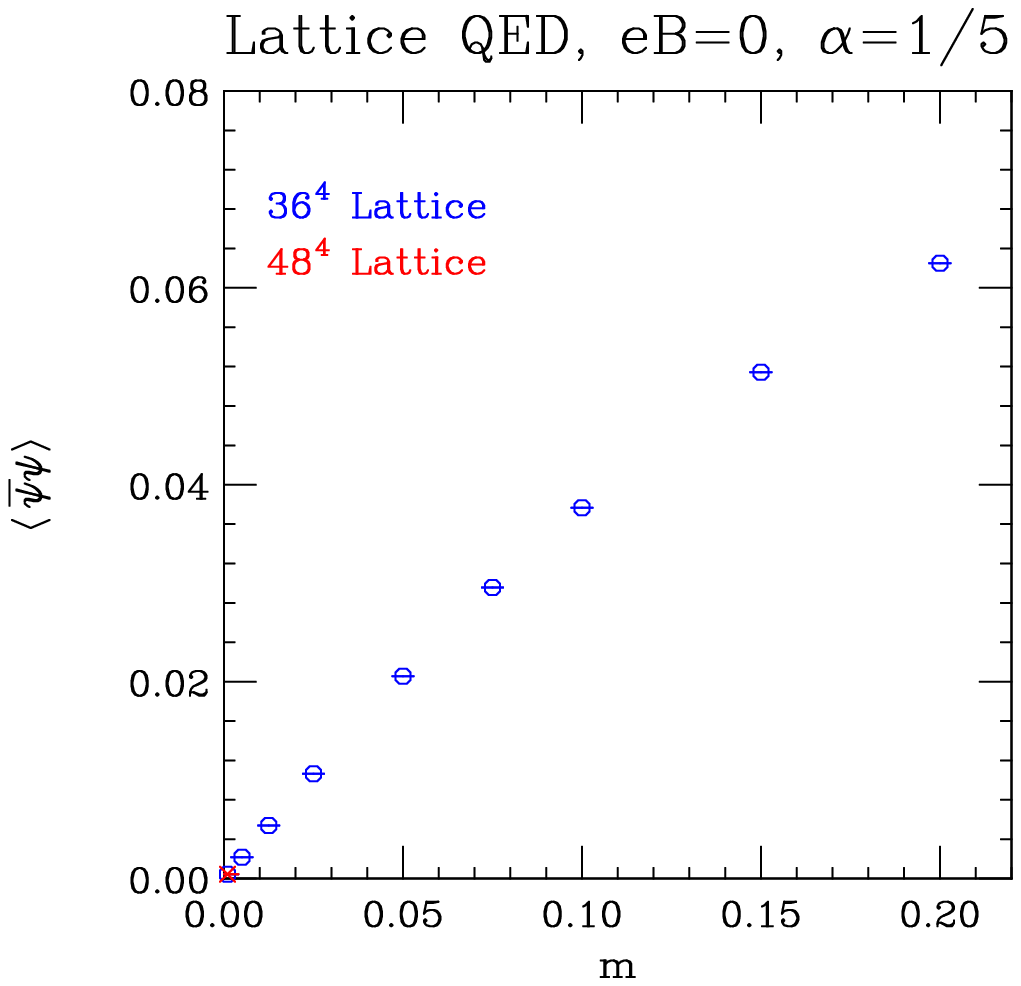}
\centerline{     }
\epsfxsize=4.0in
\epsffile{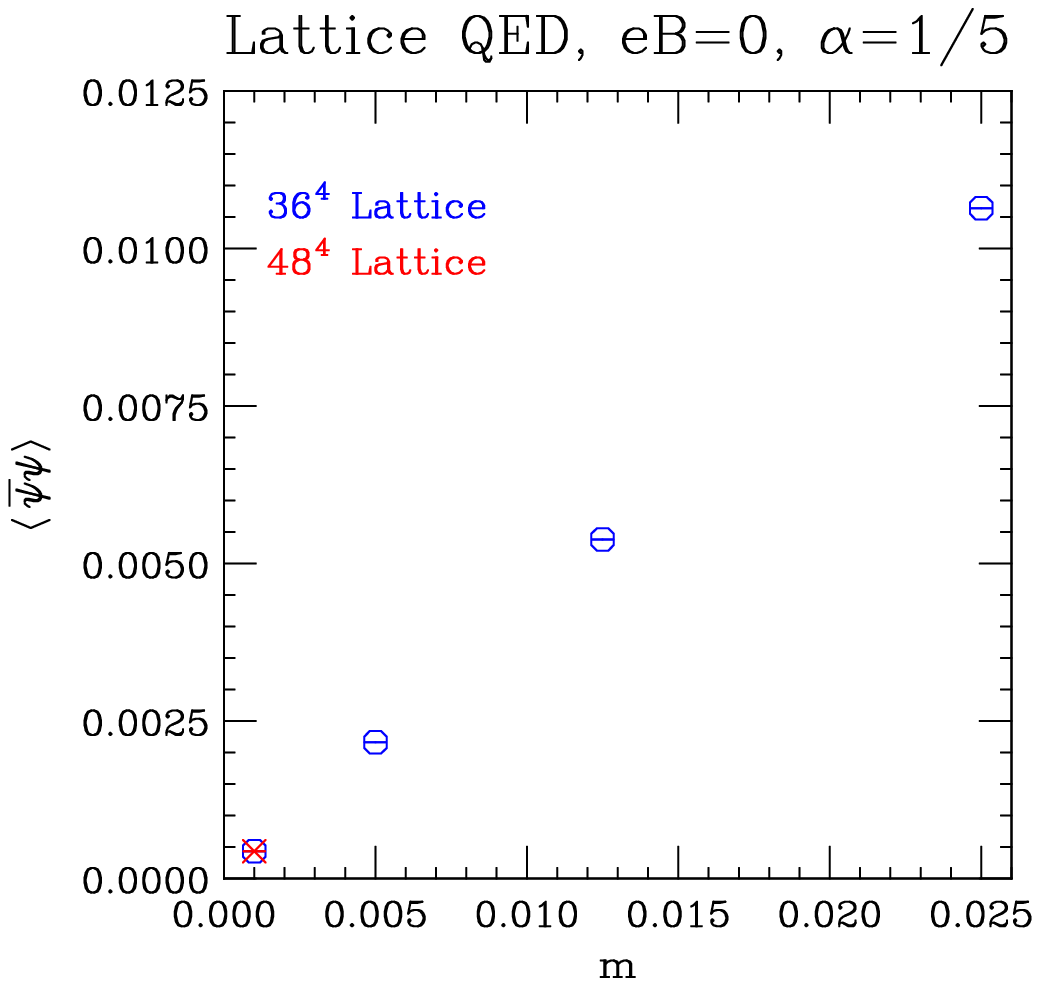}
\caption{Electron chiral condensates $\langle\bar{\psi}\psi\rangle$ as
functions of $m$ for $\alpha=0$, $eB=0$: a) full mass range, b) low mass 
region.
}
\label{fig:1/5_eB=0}
\end{figure}

Figure~\ref{fig:1/5_eB=0} shows the chiral condensate 
$\langle\bar{\psi}\psi\rangle$ as a function of $m$ on a $36^4$ lattice for
$\alpha=1/5$. This graph also shows the value of the chiral condensate
at the lowest mass, $m=0.001$, from a simulation on a larger ($48^4$) lattice,
which shows no sign of any appreciable dependence on lattice size. In fact,
$\langle\bar{\psi}\psi\rangle(m=0.001)=4.3259(7) \times 10^{-4}$ on a $36^4$
lattice compared with
$\langle\bar{\psi}\psi\rangle(m=0.001)=4.3294(7) \times 10^{-4}$ on a $48^4$
lattice. A simple linear fit to the lowest 2 masses ($m=0.001$,$m=0.005$), 
which because of the curvature of this graph should yield an upper bound to 
the value at $m=0$, predicts 
$\langle\bar{\psi}\psi\rangle_{m=0}=5(1)\times10^{-7}$, a value which 
curvature can easily lower to zero. Hence the chiral condensate at $m=0$ for 
$\alpha=1/5$ and $eB=0$ is consistent with zero.
 
To search for evidence of chiral symmetry breaking at $m=0$ catalyzed by a
strong magnetic field we perform simulations at $\alpha=1/5$ and
$eB=2\pi\times100/36^2=2\pi\times25/18^2=0.4848...$, relatively large, while 
being significantly below $eB=0.63$ above which measurements of chiral 
condensates on the lattice for free fermions in an external magnetic field 
show appreciable departures from known continuum values. Here we need to make 
measurements at $m$ values small enough that our $36^4$ lattice is too small 
to yield infinite lattice values for the chiral condensate. However, the 
assumption that for large $eB$ only the lowest Landau levels (LLL) make 
significant contributions to physics means that as long as the lattice 
projection in the $xy$ plane is considerably larger than that of the LLL, 
whose radii are $\approx 1/\sqrt{eB}$, then the chiral condensates should not 
depend on $N_x=N_y$, no matter how small $m$ becomes. For this reason we fix 
$N_x=N_y=36$ or $N_x=N_y=18$ for our simulations. On the other hand, we expect
that, at small mass, the chiral condensates will depend on the lattice extents
in the $z$ and $t$ directions.

We perform simulations at a selection of $m$ values in the range 
$0.001 \leq m \leq 0.2$, starting with a $36^4$ lattice at each mass.
At the lower masses we then increase the lattice size to $36^2\times64^2$.
Because at $m=0.025$, the increase in the chiral condensate in going from a
$36^4$ lattice to a $36^2\times64^2$ lattice is very small, we conclude that
a $36^4$ lattice would have been adequate, and that it is unnecessary to
increase the lattice sizes for any $m>0.025$. For $m=0.0125$ there is a 
small but significant increase in the chiral condensate in going from a $36^4$
lattice to a $36^2\times64^2$ lattice, from which we conclude that a $36^4$
lattice is too small, but a $36^2\times64^2$ lattice is probably adequate.
For $m=0.005$ we conclude that a $36^2\times96^2$ lattice is adequate while
for $m=0.001$ we needed the $18^2\times128^2$ lattice.  

\begin{figure}[htb]
\epsfxsize=4.0in
\epsffile{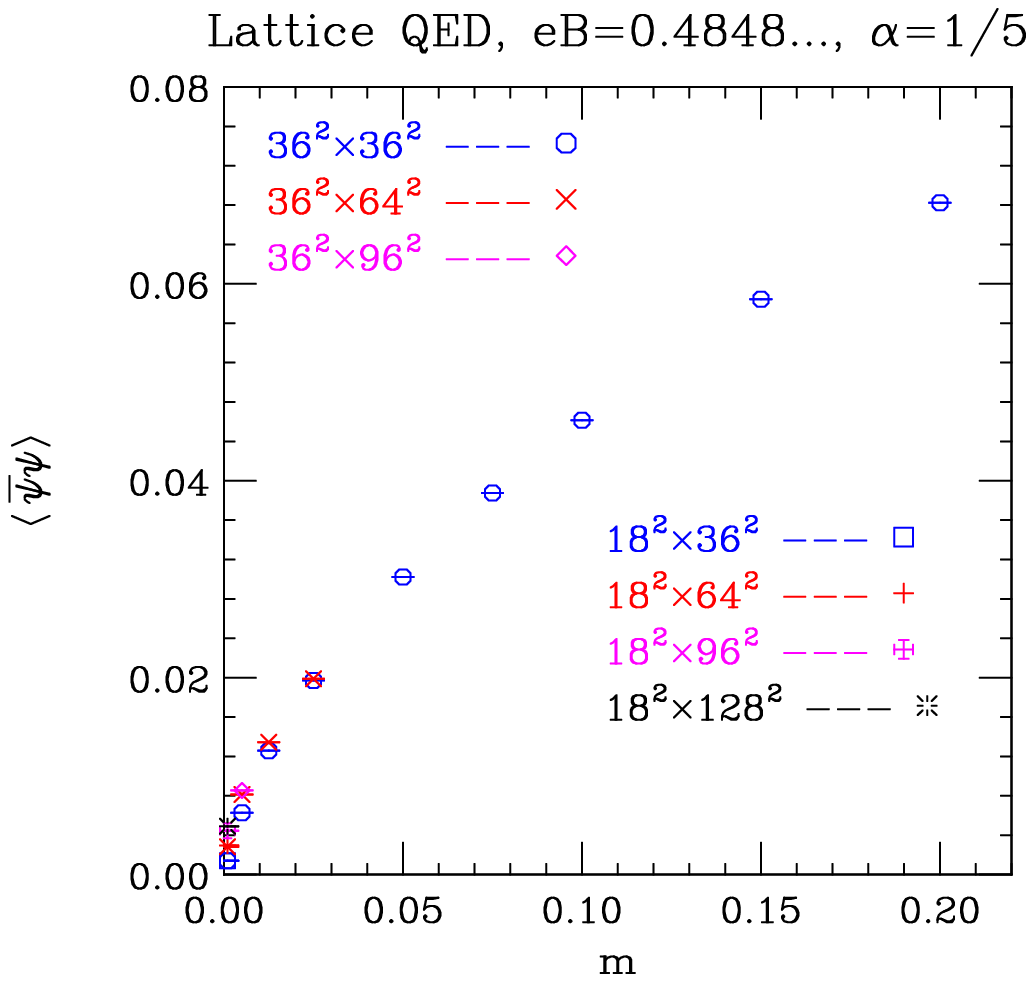}
\centerline{     }
\epsfxsize=4.0in
\epsffile{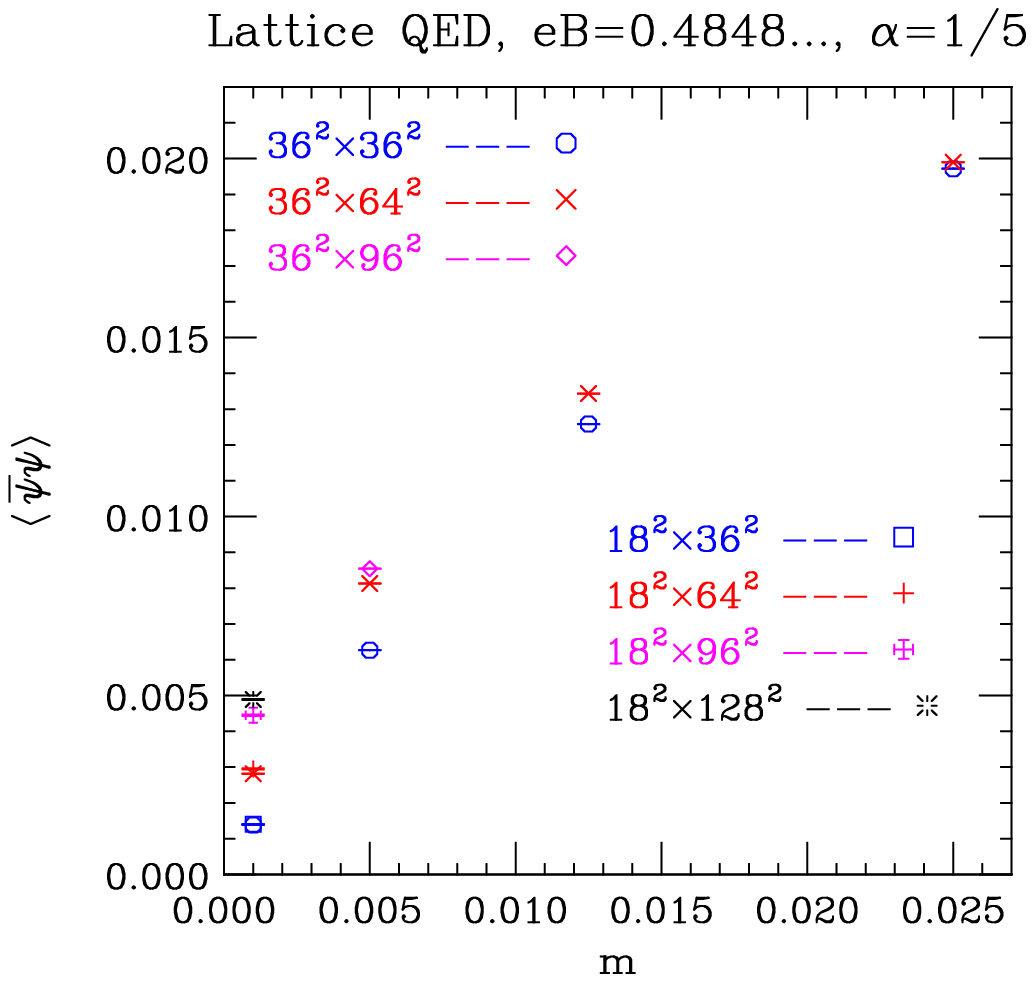}
\caption{Electron chiral condensates $\langle\bar{\psi}\psi\rangle$ as
functions of $m$ for $\alpha=0$, $eB=2\pi\times100/36^2$: a) full mass range, 
b) low mass region.}
\label{fig:1/5_eB=100}
\end{figure}

Figure~\ref{fig:1/5_eB=100} shows the chiral condensates as functions of mass
at $\alpha=1/5$ for $eB=2\pi\times100/36^2=2\pi\times25/18^2=0.4848...$. This
strongly suggests that the condensates approach a finite, non-zero limit as
$m \rightarrow 0$. From the obvious curvature of this graph (through the
uppermost point at each $m$), a straight line through the points for the 2
lightest will pass through $m=0$ at a point marking an upper bound to the
$m=0$ condensate. This value is $3.977...\times 10^{-3}$. To estimate a lower
bound, we need to fit a smooth curve to the condensates for the smallest masses
for which we have simulations. This requires choosing a functional form which
displays increasing curvature as $m \rightarrow 0$, and fitting it to the
condensates for the lowest masses. The first form we choose is:
$f(m) = a + b m + c m \log(m) + d m^2$. Fitting this to the condensates at 
$m=0.001$, $m=0.005$, $m=0.0125$ and $m=0.025$ gives values for $a,b,c,d$. The
zero mass intercept is $a=3.45...\times 10^{-3}$. Adding more points to the fit
increases the $m=0$ intercept. The curvature shows the expected behaviour.
Setting $d=0$ and fitting only the points $m=0.001$, $m=0.005$, and $m=0.0125$
raises the intercept slightly to $a=3.5097...\times 10^{-3}$, and suggests
setting $c=b$ and only using $m=0.001$, $m=0.005$ for the fit. Thus we have
reduced $f(m)$ to $f(m) = a + b m ( 1 + \log(m) )$. This lowers the
intercept slightly to $a=3.50565\times 10^{-3}$. Including the point at 
$m=0.0125$ to the fit changes the parameters $a$ and $b$ only slightly and
gives an excellent fit ($\chi^2/DOF \approx 0.7$). We now consider an 
alternative fit to $f(m) = a + b m^c$. Fitting this to the condensates at
$m=0.001$, $m=0.005$, $m=0.0125$ yields an intercept 
$a\approx3.16\times10^{-3}$. Adding the point at $m=0.025$ to the fit gives
an excellent fit ($\chi^2/DOF \approx 0.5$), lowers the intercept to
$a\approx3.14\times10^{-3}$ and makes slight changes to $b$ and $c$). In 
conclusion, a conservative estimate of the condensate at $m=0$ is that it
lies in the range:
$$
3\times10^{-3} \leq \langle\bar{\psi}\psi\rangle_{m=0} < 4\times10^{-3}
$$.

This compares with the `best' Schwinger-Dyson estimates for the chiral 
condensate at $\alpha=1/5$, $eB=0.4848...$, $N_f=1$, $m=0$, which predicts
that $\langle\bar{\psi}\psi\rangle_{m=0}\approx1.2\times10^{-4}$. Of course, a
direct comparison is not possible because the parameters in the lattice are
bare (lattice) parameters, while those in the Schwinger-Dyson estimates are 
renormalized quantities. We discuss this further in the Discussion and 
Conclusions section.

\section{Discussion and Conclusions}

We simulate lattice QED in a strong external magnetic field using the RHMC 
method developed for simulating lattice QCD. Much of our effort is aimed at
seeking evidence for chiral symmetry breaking in the limit that the lattice
(bare) mass approaches zero, catalyzed by the external magnetic field, as
predicted by less reliable truncated Schwinger-Dyson analyses. Such chiral
symmetry breaking is facilitated in part by the electrons and positrons 
preferentially occupying the lowest Landau level (LLL) with its radius
$~1/\sqrt{eB}$ due to the external magnetic field causing an effective
dimensional reduction from $3+1$ to $1+1$ for charged particles. The attractive
force due to QED then produces chiral symmetry breaking. The broken chiral
symmetry manifests itself by producing a dynamical mass for the electron and a
non-zero chiral condensate.

%Our simulations use a relatively large bare lattice fine structure constant
%$\alpha=e^2/4\pi=1/5$ and external magnetic field $eB=0.4848...$ in order for 
%us to get a measurable signal on a lattice whose extent in the plane 
%orthogonal
%to the magnetic field is $36^2$ or $18^2$. Our simulations over a range of
%bare electron masses as low as $0.001$, which require the lattice extents 
%in the direction of the field and time plane as large as $128^2$ enable us to 
%extrapolate the chiral condensate to bare mass zero with a non-zero limit.
%
%Our best estimate of the chiral condensate in the zero bare mass limit is
%$\langle\bar{\psi}\psi\rangle=3$--$4\times10^{-3}$ compared to 
%$\approx 1.2 \times 10^{-4}$ from the best Schwinger-Dyson estimates. 

We simulate Lattice QED in a constant (in space and time) external magnetic 
field $\bm{B}$, ($e\bm{B}=\bm{\nabla}\times\bm{A_{ext}}$), with staggered 
electrons, using the RHMC method. We choose the bare $\alpha=e^2/4\pi=1/5$,
$eB=2\pi\times100/36^2=0.4848...$ and a range of electron masses and lattice
sizes sufficient to extrapolate the chiral condensate 
$\langle\bar{\psi}\psi\rangle$ to $m=0$. We estimate
$\langle\bar{\psi}\psi\rangle=3$--$4\times10^{-3}$ at m=0, definitely non zero,
indicating that chiral symmetry is broken. Note that the parameters and fields
in the action are bare lattice quantities. This compares with the renormalized 
condensate of $\approx 1.2 \times 10^{-4}$ from the best Schwinger-Dyson 
estimates. Since the lowest non-trivial contributions to the renormalization
condensates evaluated at $p^2=m_{dyn}^2$ are typically 
${\cal O}(\alpha\log(\Lambda^2/m_{dyn}^2))\approx1$--$3$ where 
$\Lambda\approx\pi$ is the momentum cutoff on the lattice, renormalization
cannot be ignored. This contrasts with our simulations with $\alpha=1/137$,
$m=0.1,0.2$, where the renormalization constants are evaluated at $p^2=m^2$,
so the lowest non-trivial order corrections are typically 
${\cal O}(\alpha\log(\Lambda^2/m^2)\approx0.05,0.04)$ and can be ignored if
one can tolerate errors of a few percent (in actual fact for small external
magnetic fields the errors are somewhat smaller than this). There are two ways
that one might check the consistency between the Schwinger-Dyson approach
and the Lattice QED simulations. The first is to follow the methods developed
for renormalizing lattice QCD. The dimensional reduction from $3+1$ to $1+1$
dimensions caused by the external magnetic field adds extra complications.
The second approach is to repeat the Schwinger-Dyson analysis on the lattice 
action based on methods developed for lattice perturbation theory, in terms of
bare parameters. Of course, it would be best to use both methods.

Experience with lattice QCD indicates that we should expect large taste 
breaking effects from the use of staggered fermions (and, in particular, rooted
staggered fermions). See for example the review article \cite{MILC:2009mpl},
and its guide to the literature. In lattice QCD these effects are, to lowest 
order, ${\cal O} (a^2\Lambda_{QCD}^2)$ where $a$ is the lattice spacing. For 
lattice QED in an external magnetic field these effects should be, to lowest 
order ${\cal O} (a^2 eB)$ or, since we have chosen $a=1$, ${\cal O} (eB)$. 
Hence we should be able to reduce taste breaking by reducing the external 
magnetic field. We are therefore repeating our simulations with a smaller 
magnetic field. Another way one can see that this should decrease errors is to
note that with a smaller magnetic field, the lowest Landau level covers more 
lattice sites so that discretization errors should be reduced. With the smaller
magnetic field, we should be able to test that the chiral condensate scales
as $(eB)^{3/2}$ as predicted.

It would be interesting to apply lattice simulations to QED with $N_f=2$ 
flavours in an external magnetic field, where such chiral symmetry breaking
would imply spontaneous dynamical breaking of flavour symmetry with associated
Goldstone bosons. Note that these Goldstone bosons are uncharged and hence
reside in $3+1$ dimensions and therefore spontaneous breaking of continuous
chiral symmetry is allowed despite the electrons being restricted to $1+1$
dimensions.

One of the first calculations planned for stored configurations is to measure
the effect that QED in an external magnetic field has on the coulomb field of
a point charge in said magnetic field. It has been predicted that the coulomb
field will be partially screened and distorted by the presence of the external
magnetic field \cite{Shabad:2007xu,Shabad:2007zu,Sadooghi:2007ys,%
Machet:2010yg}. This effect can and will be measured on our stored 
configurations at $\alpha=1/137$ where renormalization can be safely ignored,
with expected errors of at most a few percent. Such an analysis will be
performed using Wilson loops, possibly with smearing.

We plan to calculate electron propagators on our stored configurations with
$\alpha=1/5$, $eB=0.4848...$ on lattice sizes from $36^2\times64^2$ to
$18^2\times128$ to extract the electron mass extrapolated to $m=0$. When
configurations from our current and ongoing simulations at $eB=0.1163...$
become available we will then be able to check that $m_{dyn}\propto\sqrt{eB}$,
and extract the electron wave-function renormalization constant. We will also
attempt to calculate the photon propagator, although this will be considerably
more difficult, since its falloff with separation of its endpoints will be
statistics limited.

The one parameter left to calculate will be the renormalized coupling constant
which, unless this can be calculated perturbatively, might well require 
additional simulations.

Of at least as much interest is the behaviour of QED in an external electric
field, which exhibits the Sauter-Schwinger effect. As pointed out earlier,
this has a complex action and traditional lattice simulations which rely on
importance sampling cannot be used. A good starting point is to simulate
lattice QED with both external electric and magnetic fields where these fields
are obtained by a boost from a pure external magnetic field, since the physics
should be the same as that for this magnetic field alone. One method we might
try is a Complex Langevin (CLE) simulation. This has some chance of succeeding
for QED in external electric fields or electric and magnetic fields, whereas
it failed for QCD in a quark-number chemical potential, because for pure 
non-compact $U(1)$ lattice gauge theory, which is a free field theory and
hence a collection of harmonic oscillators, the real trajectory is an 
attractive fixed-point of the CLE, while for pure compact $SU(3)$ lattice gauge
theory it is a repulsive fixed point. We will first need to check if the real 
trajectory remains an attractive fixed point for the $U(1)$ lattice gauge 
theory CLE simulation when the electron fields are included making it lattice 
QED, but without the external electric fields. If so, we plan to try CLE when 
the external electric or electric and magnetic fields are added. 

\acknowledgments

DKS's research is supported in part by U.S. Department of Energy, Division of
High Energy Physics, under Contract No. DE-AC02-06CH11357. The high performance
computing was provided by the LCRC at Argonne National Laboratory on their
Bebop cluster. Access to Stampede-2 at TACC, Expanse at UCSD and Bridges-2 at
PSC was provided under an XSEDE/ACCESS allocation. Time on Cori and Perlmutter
at NERSC was provided through an ERCAP allocation and from early user access to
Perlmutter during its pre-acceptance period. DKS thanks G.~T.~Bodwin for 
inciteful discussions, while JBK would like to thank V.~Yakimenko for 
discussions which helped inspire this project, and I.~A.~Shovkovy for helpful 
discourse on magnetic catalysis.

\end{document}